\documentclass[12pt]{article}
\usepackage{color}
\usepackage{latexsym}
\usepackage{epsfig,amssymb,euscript,mathrsfs}
\usepackage{amsmath}
\newsavebox{\ns}
\newsavebox{\dbrane}
\newsavebox{\dbshort}

\def\be{\begin{equation}}
\def\ee{\end{equation}}
\def\bea{\begin{eqnarray}}
\def\eea{\end{eqnarray}}

\newcommand\R{\mathbb{R}}
\newcommand\Z{\mathbb{Z}}

\newcommand\diff{\mathrm{d}}

\newcommand{\dd}{\mathrm{d}}
\newcommand{\me}{\mathrm{e}}
\newcommand{\ii}{\mathrm{i}}

\newcommand{\ex}{\mathrm{e}}

\newlength{\sswidth}

\newcommand{\gr}{\mathrm{grav}}
  
\newcommand{\bd}{\mathrm{bdry}}
\newcommand{\ct}{\mathrm{ct}}
\newcommand{\free}{\mathcal{F}}
\newcommand{\s}{s}  

\numberwithin{equation}{section}       % equation numbers in each section

\begin{document}

\begin{titlepage}

\begin{center}

\today

\vskip 2.3 cm 

{\Large \bf The gravity dual of supersymmetric gauge theories
\vskip 5mm
on a biaxially squashed three-sphere}

\vskip 2 cm

{Dario Martelli$^1$ and James Sparks$^2$\\}

\vskip 1cm

$^1$\textit{Department of Mathematics, King's College, London, \\
The Strand, London WC2R 2LS,  United Kingdom\\}

\vskip 0.8cm

$^2$\textit{Mathematical Institute, University of Oxford,\\
24-29 St Giles', Oxford OX1 3LB, United Kingdom\\}

\end{center}

\vskip 2 cm

\begin{abstract}
\noindent We present the gravity dual to a class of three-dimensional $\mathcal{N}=2$ supersymmetric gauge theories on 
a biaxially squashed three-sphere, with a non-trivial background gauge field. This is described by a 1/2 BPS Euclidean solution of four-dimensional $\mathcal{N}=2$ gauged 
supergravity, consisting of a Taub-NUT-AdS metric with a non-trivial instanton for the graviphoton field. 
The holographic free energy of this solution agrees precisely with the large $N$ limit of the free energy obtained from the localized partition function 
of a class of Chern-Simons quiver gauge theories. We also discuss a different supersymmetric 
solution, whose boundary is a biaxially squashed Lens space $S^3/\Z_2$ with a topologically non-trivial background gauge field. 
This metric is of Eguchi-Hanson-AdS type, although it is not Einstein, and has a single unit of gauge field flux through the $S^2$ cycle.
 \end{abstract}

\end{titlepage}

\pagestyle{plain}
\setcounter{page}{1}
\newcounter{bean}
\baselineskip18pt

\section{Introduction}

Supersymmetric gauge theories on compact curved backgrounds are interesting for various reasons. For example, supersymmetry may be combined with 
localization techniques, allowing one to perform a variety of exact computations in strongly coupled field theories. The authors of
\cite{Hama:2011ea} presented a construction of ${\cal N}=2$ supersymmetric gauge theories in three dimensions in the background of a $U(1)\times U(1)$-invariant
squashed three-sphere and R-symmetry gauge field. 
The gravity dual of this construction was recently given in  \cite{Martelli:2011fu}. It consists of a 1/4 BPS Euclidean solution of four-dimensional $\mathcal{N}=2$ gauged 
supergravity, which in turn may be uplifted to a supersymmetric solution of eleven-dimensional supergravity. In particular, the bulk metric in 
\cite{Martelli:2011fu} is simply AdS$_4$,  and the graviphoton field is an instanton with (anti)-self-dual field strength. The asymptotic metric and gauge field then reduce to the
background considered in \cite{Hama:2011ea}.

The purpose of this letter is to present the gravity dual to a different field theory construction, obtained recently in 
\cite{Imamura:2011wg}. In  this reference the authors have constructed three-dimensional ${\cal N}=2$ supersymmetric gauge theories 
in the background of the $SU(2)\times U(1)$-invariant squashed three-sphere (which we refer to as  \emph{biaxially} squashed) 
and a non-trivial background $U(1)$  gauge field, and have computed the corresponding partition functions using localization. 
Differently from a %superficially 
similar construction discussed briefly in \cite{Hama:2011ea}, this partition function depends non-trivially on the squashing parameter.
As we will see, the gravity dual to this set-up will have some distinct features with respect to the solution in \cite{Martelli:2011fu}. In particular, the metric 
is not simply AdS$_4$, although it will again be an Einstein metric, and there is  a self-dual graviphoton. 

The plan of the rest of this paper is as follows. In section \ref{revsect} we review the construction 
of \cite{Imamura:2011wg}. In section \ref{nuts} we discuss the gravity dual. In section \ref{bolty} we describe a different 
supersymmetric solution, consisting of a non-Einstein metric and a non-instantonic graviphoton field. 
Section \ref{discussione} concludes.

\section{Supersymmetric gauge theories on the  biaxially squashed $S^3$}

\label{revsect}

In the construction of  \cite{Imamura:2011wg}
the metric on the three-sphere is, up to an irrelevant  overall factor, given by
\bea
\diff s^2_3 & = & \sigma_1^2 + \sigma_2^2 + \frac{1}{v^2}\sigma_3^2~,
\label{fametric}
\eea 
where $\sigma_i$ are the standard $SU(2)$ left-invariant one-forms on $S^3$, defined as $\ii \sigma_i \tau_i = -2 \mathtt{g}^{-1}\diff \mathtt{g}$, where 
$\tau_i$ denote the Pauli matrices and  $\mathtt{g}\in SU(2)$. The background $U(1)$ gauge field reads
\bea
A^{(3)} & = & \frac{\sqrt{v^2-1}}{2v^2}\sigma_3 ~,
\label{newgau}
\eea
and the spinors in the supersymmetry transformations obey the equation (setting the radius $r=2$ in \cite{Imamura:2011wg})
\bea
\nabla_\alpha^{(3)}\chi- \frac{\ii}{4v}\gamma_\alpha \chi - A^{(3)}_\beta \gamma_\alpha{}^\beta \chi & = & 0~,
\label{newks}
\eea
where $\nabla_\alpha^{(3)}$,   $\alpha=1,2,3$, is the spinor covariant derivative constructed from the metric (\ref{fametric}),  and 
$\gamma_\alpha$ generate $\mathrm{Cliff}(3,0)$. There are \emph{two} linearly independent solutions to 
(\ref{newks}), transforming as a doublet under $SU(2)$, whose explicit form is given in \cite{Imamura:2011wg}. This will be important for identifying the gravity dual. 

 In  \cite{Imamura:2011wg} the authors 
constructed Chern-Simons, Yang-Mills, and matter Lagrangians for  the ${\cal N}=2$ 
vector multiplets $V=(\mathscr{A}_\alpha,\sigma,\lambda,D)$ and chiral multiplets $\Phi = (\phi,\psi,F)$, 
in the background of the metric  (\ref{fametric}) and R-symmetry  gauge field (\ref{newgau}).
These  are 
invariant under a set of supersymmetry transformations, provided the spinorial parameters obey the equation (\ref{newks}). 
The supersymmetric completion of the Chern-Simons Lagrangian contains new terms, in addition to those appearing in flat space, 
proportional to $\sigma^2$ and $\sigma A^{(3)}\wedge \diff \mathscr{A}$ (\emph{cf}. eq. (32) 
of \cite{Imamura:2011wg}). The Yang-Mills and matter Lagrangians are total supersymmetry variations (\emph{cf}. eq. (31) of \cite{Imamura:2011wg})
and therefore can be used to compute the partition function using localization. In particular, the partition function 
localizes on supersymmetric configurations obeying
\bea
\mathscr{A}_\alpha \ = \  D \ = \ 0 ~, \qquad \sigma  \ = \ u \ =  \ \mathrm{constant}~,
\label{newloc}
\eea
with the matter fields all being zero.
Notice that although $D=0$, the Chern-Simons Lagrangian 
is non-zero because of the new  term proportional to $\sigma^2$, and therefore it contributes classically to the localized partition function, as in previous constructions. 
The Yang-Mills and matter terms contribute one-loop determinants from the Gaussian 
integration about the the classical solutions (\ref{newloc}). The final partition function may be expressed again in terms of double sine functions $s_b(z)$, and 
for a $U(N)$ gauge theory at Chern-Simons level $k\in\Z$ reads
\bea\label{partition}
Z & = & \int\prod_{\mathrm{Cartan}} \diff u\,  \exp\left(\frac{\ii \pi k}{v^2}\mathrm{Tr}\, u^2\right)\frac{\prod_{\mathrm{Roots}\, \alpha} s_b\left(\frac{\alpha(u)-\ii}{v}\right)}{
\prod_{\mathrm{Chirals}, \, \mathrm{rep}\, \mathcal{R}_a} \prod_{\rho\in \mathcal{R}_a} s_b\left(\frac{\rho(u)-\ii(1-\Delta_a)}{v}\right)}~,
\eea 
where $b=(1+\ii\sqrt{v^2-1})/v$.
The exponential term is the classical contribution from the Chern-Simons Lagrangian, evaluated on (\ref{newloc}); the numerator is the one-loop
vector multiplet determinant and
involves a product over the roots $\alpha$ of the gauge group $G$; while the denominator is the one-loop
matter determinant and involves a product over chiral fields of R-charge $\Delta_a$ in representations $\mathcal{R}_a$, with $\rho$ running 
over weights in the weight-space decomposition of $\mathcal{R}_a$. 
Following \cite{Martelli:2011qj}, one can easily extract  the large $N$ behaviour of this partition function for a 
class of non-chiral ${\cal N}=2$ quiver Chern-Simons-matter theories. The calculation was done in \cite{Imamura:2011wg}, and the result is that the leading contribution to the free energy (defined as $\free=-\log Z$) is given by
\bea
\free_v & = & \frac{1}{v^2}\, \free_{v=1}~,
\label{newfree}
\eea
and thus depends very simply on the squashing parameter $v$. In the next section we will present the supergravity dual to this construction, in particular 
showing that the holographic free energy precisely agrees with the field theory result (\ref{newfree}).

\section{The gravity dual}\label{nuts}

As anticipated in \cite{Martelli:2011fu}, we will show that the gravity dual to the set-up described in the previous section is
a supersymmetric solution of $d=4$, $\mathcal{N}=2$ gauged supergravity. 
In \emph{Lorentzian} signature, the bosonic part of the action is given by
\bea\label{4dSUGRA}
S_{\mathrm{Lorentzian}} &=& \frac{1}{16\pi G_4}\int \diff^4x\sqrt{-\det g_{\mu\nu}}\left[R + 6g^2 - (F^L)^2\right]~.
\eea 
Here $R$ denotes the Ricci scalar of the four-dimensional metric $g_{\mu\nu}$, and the cosmological constant 
is given by $\Lambda=-3g^2$. The graviphoton is an Abelian gauge field $A^L$ with field strength 
$F^L=\diff A^L$; here the superscript $L$ emphasizes that this is a Lorentzian signature object.
A solution to the equations of motion derived from (\ref{4dSUGRA})
 is supersymmetric if there is a non-trivial spinor $\epsilon$
satisfying the Killing spinor equation
\bea\label{LKSE}
\left[ \nabla_\mu + \tfrac{1}{2} g \Gamma_\mu - \ii g A^L_\mu + \tfrac{\ii}{4} F^L_{\nu\rho} \Gamma^{\nu\rho} \Gamma_\mu \right] \epsilon &=& 0~.
\eea
Here $\Gamma_\mu$, $\mu=0,1,2,3$, generate the Clifford algebra $\mathrm{Cliff}(1,3)$, so 
$\{\Gamma_\mu,\Gamma_\nu\}=2g_{\mu\nu}$.

Since the background of \cite{Imamura:2011wg} preserves half of the maximal supersymmetry in three dimensions, 
we should seek a \emph{1/2 BPS Euclidean} solution of $d=4$, $\mathcal{N}=2$ gauged supergravity, 
whose metric has as conformal boundary the biaxially squashed metric on $S^3$ (\ref{fametric}),
and whose background $U(1)$ gauge field restricted to this asymptotic boundary  
reduces to (\ref{newgau}). This very strongly suggests that the appropriate solution 
is a Euclideanized version of the 1/2 BPS Reissner-Nordstr\"om-Taub-NUT-AdS solution discussed 
 in \cite{AlonsoAlberca:2000cs}. 

We will first present this Euclidean solution, and then discuss the Wick rotation 
that leads to it. The metric reads
\bea
\diff s^2_4 &= &  \frac{r^2-\s^2}{\Omega(r)}\diff r^2 +
 (r^2-\s^2)(\sigma_1^2+\sigma_2^2) +\frac{4\s^2\Omega(r)}{r^2-\s^2}\sigma_3^2 ~,
\label{TNAdS}
\eea
where 
\bea
\Omega (r) & = &  (\s - r)^2 \left[1 + g^2 (r - \s) (r+3 \s )\right]~,
\label{finalome}
\eea
and $\s$ is the NUT parameter.\footnote{This is denoted $N$ in  \cite{AlonsoAlberca:2000cs}.} 
The  $SU(2)$ left-invariant one-forms  $\sigma_i$ may be written in terms of angular variables
as
\bea
\sigma_1 + \ii \sigma_2 \, =\, \me^{-\ii \psi}  ( \diff \theta + \ii \sin \theta \diff \varphi) ~,\qquad \sigma_3 \, =\, \diff \psi + \cos \theta \diff \varphi~.
\eea
The graviphoton field is
\bea
A & = &  s\frac{r-s}{r+s}\sqrt{1-4g^2s^2}  \, \sigma_3~.
\label{gaugefield}
\eea
In the orthonormal frame
\bea\label{frame}
e_1 &=& \sqrt{r^2-s^2}\, \sigma_1~, \qquad \ \ \  \, e_2 \ = \  \sqrt{r^2-s^2}\, \sigma_2~,\nonumber\\
e_3  & = & 2s \sqrt{\frac{\Omega(r)}{r^2-s^2}}\sigma_3~, \qquad e_4 \ = \ \sqrt{\frac{r^2-s^2}{\Omega(r)}}\diff r~,
\eea
the curvature may be written as
\bea
F &=& \diff A \ = \ - \frac{\s\sqrt{1-4g^2\s^2}}{(r+\s)^2}\left(e_{12}+e_{34}\right)~.
\label{instantfield}
\eea
Thus the gauge field is an \emph{instanton}, as in the solution discussed in \cite{Martelli:2011fu}.
In particular, with our choice of orientation the curvature is self-dual, and the on-shell gauge field action is finite. 
Since the stress-energy tensor of an instanton vanishes, the metric (\ref{TNAdS})  is accordingly  an Einstein metric.
However, differently from the solution in  \cite{Martelli:2011fu},
one can check that this metric is \emph{not} locally AdS$_4$.
It is in fact a Euclidean version of the well-known Taub-NUT-AdS metric, with a special value of the mass parameter. This metric is 
locally asymptotically AdS$_4$, and therefore it can be interpreted holographically  \cite{Chamblin:1998pz}. 
Notice that for $|s|\leq 1/(2g)$ the gauge field (\ref{gaugefield}) is \emph{real}, while for $|s|> 1/(2g)$ it is \emph{purely imaginary};
the intermediate case with  $|s|=1/(2g)$ has vanishing gauge field instanton and the metric 
reduces to Euclidean AdS$_4$.

For large $r$  the metric becomes
\bea\label{approxmetric}
\diff s^2_4 & \approx & \frac{\diff r^2}{g^2 r^2}+ r^2\left[\sigma_1^2+\sigma_2^2+{4g^2\s^2}\sigma_3^2\right]~,
\eea
while to leading order the gauge field reduces to 
\bea\label{A3}
A & \approx  & A^{(3)} \ \equiv \   s\sqrt{1-4g^2s^2}  \, \sigma_3~.
\eea
We see that the conformal boundary may be identified \emph{precisely} with the metric (\ref{fametric}), and the background gauge field with (\ref{newgau}),
by setting $s = \tfrac{1}{2gv}$. Recall here  that in order to uplift to eleven-dimensional supergravity one should also set $g=1$ \cite{Martelli:2011fu}. 
Notice that when $|v|=1$ the boundary metric reduces to the round metric on $S^3$, and the background gauge field vanishes. 
Correspondingly, in the bulk the instanton field vanishes, and the metric becomes AdS$_4$.  

\subsection*{Wick rotation and regularity}

Let us discuss briefly  how this solution was obtained. The reader not interested in these details may safely jump to the discussion of the Killing spinors and the holographic free energy. 

As we are interested in a 1/2 BPS solution, we 
may begin by appropriately Wick rotating  the solution (2.1), (2.4) of  \cite{AlonsoAlberca:2000cs}. We take their parameter 
$\aleph=+1$, so as to obtain a biaxially squashed $S^3$ as constant $r$ surface. The Wick rotation may then be taken to be 
$t\rightarrow \ii\tau$, $N\rightarrow \ii s$, $Q\rightarrow \ii Q$, together with a change in sign of the metric. This leads to the following 
metric and gauge field
\bea
\diff s^2_4 &=&  \frac{r^2-\s^2}{\Omega(r)}\diff r^2 + (r^2-\s^2)(\diff\theta^2+\sin^2\theta\diff\varphi^2) + \frac{\Omega(r)}{r^2-\s^2}\left(\diff \tau+ 2 \s \cos\theta\diff\varphi\right)^2~,\nonumber
\\\label{intermedi}
A^L &=& \frac{\s P - Qr}{r^2-\s^2}\diff\tau + \frac{P(r^2+\s^2)-2\s Qr}{r^2-\s^2}\cos\theta\diff\varphi~,
\eea
where 
\bea
\Omega(r) &=& g^2 (r^2-\s^2)^2 + (1-4g^2\s^2)(r^2+\s^2)-2Mr + (P^2-Q^2)~.
\label{genom}
\eea
This depends on the parameters $s,g,M,P,Q$. Notice we have kept a Lorentzian superscript on $A^L$ in (\ref{intermedi}) -- the reason for this will become 
clear momentarily. 

For the 1/2 BPS solution of interest, 
the Euclideanized BPS equations  of \cite{AlonsoAlberca:2000cs} imply that 
\bea
M^2 &=& (1-4g^2\s^2)\left[\s^2(1-4g^2\s^2)+P^2-Q^2\right]~,\nonumber\\
\s^2 P(1-4g^2\s^2) &=& sMQ - P(P^2-Q^2)~,\label{BPS}
\eea
and the corresponding 1/2 BPS solution then depends on only \emph{two} parameters. We take these to be 
$s$ and $Q$, with 
\bea
P & =& \ii\s\sqrt{1-4g^2\s^2}~,\qquad M \  =\  -\ii Q\sqrt{1-4g^2\s^2}~,
\label{jimllfixit}
\eea
then solving (\ref{BPS}). The factors of $\ii$ in (\ref{jimllfixit}) may look problematic, but there are (at least) two different 
ways of obtaining real solutions. We require $s$ and $M$ to be real in order that the metric in (\ref{intermedi}) is real. 
If $|s| \leq 1/(2g)$ then $P$ and $Q$ will be purely imaginary, and we may write $P=\ii p$, $Q=-\ii q$ 
to obtain the \emph{real} gauge field
\bea\label{Wiki}
A &\equiv & -\ii A^L \ = \ \frac{\s p + qr}{r^2-\s^2}\diff\tau + \frac{p(r^2+\s^2)+2\s qr}{r^2-\s^2}\cos\theta\diff\varphi~.
\eea

Redefining  $\tau = 2\s \psi$, in terms of standard Euler angles $(\theta,\varphi,\psi)$  notice that the metric (\ref{intermedi}) takes the form
presented in  (\ref{TNAdS}), albeit with a more general form of the function $\Omega (r)$, given by 
(\ref{genom}) and (\ref{jimllfixit}). That (\ref{TNAdS}) has only one free parameter $s$, and not the two we have above, 
follows from imposing \emph{regularity} of the Euclidean metric. 
At any fixed $r>s$ that is not a root of $\Omega(r)$, we obtain a smooth biaxially squashed $S^3$ metric. 
In order to obtain a complete metric, the space must ``close off'' at the largest root $r_0$ of $\Omega(r)$, so that $\Omega(r_0)=0$. 
More precisely, if $r_0>s$ this should be a single root, while if $r_0=s$ the metric will be regular only if $r_0=s$ is a double root of $\Omega(r)$. 
We shall return to the former case in section \ref{bolty}, here focussing on the case $r_0=s$. The condition $\Omega(r_0=s)=0$ immediately fixes
\bea
q &=& -\s \sqrt{1-4g^2\s^2}~,
\eea
so that now (see also \cite{Emparan:1999pm})
\bea
p &=& \s\sqrt{1-4g^2\s^2} \ = \ -q~, \qquad  M \ = \ \s(1-4g^2\s^2)~.
\eea
It is then in fact automatic that $r=\s$ is a double root of $\Omega$.

In conclusion, we end up with the metric (\ref{TNAdS}), with $\Omega (r)$ given in (\ref{finalome}),
and gauge field  (\ref{gaugefield}). The gauge field  is manifestly non-singular and one can check that the 
metric indeed  smoothly closes off at $r=\s$, giving the topology $M_4=\R^4$.

\subsection*{Killing spinors}

In this subsection we briefly discuss the supersymmetry of the Euclidean solution 
 (\ref{TNAdS}), (\ref{gaugefield}), in particular reproducing the three-dimensional 
 spinor equation (\ref{newks}) asymptotically.
  
  In Lorentzian signature the Killing spinor equation 
 is (\ref{LKSE}). However, in Wick rotating we have introduced a factor of $\ii$ into the gauge field in (\ref{Wiki}), so that $A^L=\ii A$.
 Thus the appropriate Killing spinor equation to solve in this case is
\bea\label{funnyKSE}
\left[ \nabla_\mu + \tfrac{1}{2} g \Gamma_\mu +  g A_\mu - \tfrac{1}{4} F_{\nu\rho} \Gamma^{\nu\rho} \Gamma_\mu \right] \epsilon &=& 0~.
\eea
This possibility of Wick rotating the gauge field (or not) was also discussed in \cite{Dunajski:2010uv}. 
In particular, the authors of \cite{Dunajski:2010uv} pointed out that any Euclidean solution 
with a real gauge field that solves (\ref{funnyKSE}) will automatically be 1/2 BPS. 
The reason is simple: if $\epsilon$ solves (\ref{funnyKSE}), then so does its 
conjugate $\epsilon^c$. We shall see this explicitly below.

We introduce the following representation for the generators of Cliff$(4,0)$
\bea
\hat{\Gamma}_4 & =&  \begin{pmatrix} 0 & \ii \mathbb{I}_2 \\ -\ii \mathbb{I}_2 & 0 \end{pmatrix}~, ~~~~~~~~
\hat{\Gamma}_\alpha  \ =  \ \begin{pmatrix} 0 & \tau_\alpha \\ \tau_\alpha & 0 \end{pmatrix} ~,
\eea
where $\alpha \in 1,2,3$, $\tau_\alpha$ are the Pauli matrices, and hats denote tangent space quantities. 
Decomposing the Dirac spinor $\epsilon$ into positive and negative chirality parts as
\bea
\epsilon &=& \left(\begin{array}{c} \epsilon_+\\ \epsilon_-\end{array}\right)~,
\eea
where $\epsilon_\pm$ are two-component spinors, it is then straightforward, but tedious, to verify that in the orthonormal frame (\ref{frame})
\bea\label{spinors}
\epsilon_+ &=& \left(\begin{array}{c} \lambda(r)\chi_+\\ \lambda^*(r)\chi_-\end{array}\right)~, \qquad \epsilon_- \ = \ \ii \sqrt{\frac{r-\s}{r+\s}}\left(\begin{array}{c} 
\lambda^*(r)\chi_+ \\ \lambda(r)\chi_-\end{array}\right)~, 
\eea
is the general solution to the $\mu=r$ component of (\ref{funnyKSE}), where $\chi_\pm$ are independent of $r$
and we have defined
\bea
\lambda(r) & \equiv & \left(g(r+\s)-\ii \sqrt{1-4g^2\s^2}\right)^{1/2}~.
\eea
If we now define the charge conjugate spinor $\epsilon^c \equiv B\epsilon^*$, where 
$B$ is the charge conjugation matrix defined in \cite{Martelli:2011fu},
then it is straightforward to see that taking the conjugate $\epsilon\rightarrow \epsilon^c$ simply 
maps $\chi_+\rightarrow -\chi_-^*$, $\chi_-\rightarrow\chi_+^*$.

Let us analyze the large $r$ asymptotics of the Killing spinor equation (\ref{funnyKSE}), and its solutions (\ref{spinors}). We begin by expanding
\bea
\epsilon_+ &=& \sqrt{g}r^{1/2}\left[\mathbb{I}_2+\left(\frac{\s}{2}\mathbb{I}_2-\frac{\ii}{2g}\sqrt{1-4g^2\s^2}
\tau_3\right)r^{-1}+\mathcal{O}(r^{-2})\right]\chi~,\nonumber\\ \label{spinorexpand}
\epsilon_- &=& \ii\sqrt{g}r^{1/2}\left[\mathbb{I}_2-\left(\frac{\s}{2}\mathbb{I}_2-\frac{\ii}{2g}\sqrt{1-4g^2\s^2}
\tau_3\right)r^{-1}+\mathcal{O}(r^{-2})\right]\chi~,
\eea
where we have defined the $r$-independent two-component spinor
\bea\label{chispinor}
\chi & \equiv & \left(\begin{array}{c}\chi_+\\ \chi_-\end{array}\right)~.
\eea
We then write the asymptotic expansion of the metric as 
\bea
\dd s^2_4 &=& \frac{\diff r^2}{g^2 r^2}\left[1+\mathcal{O}(r^{-2})\right]+ \frac{r^2}{g^2}\left[\diff s^2_3+\mathcal{O}(r^{-2})\right]~,\\\label{threemetric}
\diff s^2_3 &\equiv & g^2\left[\sigma_1^2+\sigma_2^2+{4g^2\s^2}\sigma_3^2\right]~.
\eea
It is then straightforward to extract the coefficient of $r^{1/2}$ in the Killing spinor equation (\ref{funnyKSE}). 
One finds that the positive and negative chirality projections lead to the \emph{same} equation 
for $\chi$, namely
\bea\label{3dkse}
\nabla_\alpha^{(3)}\chi+ gA^{(3)}_\alpha \chi - \frac{\ii \s}{2}\gamma_\alpha \chi - \frac{1}{2g}\sqrt{1-4g^2\s^2}\gamma_\alpha\tau_3 \chi & = & 0~,
\eea
where $\nabla^{(3)}$ denotes the spin connection for the three-metric (\ref{threemetric}), 
and $A^{(3)}$ is defined in (\ref{A3}). Using the explicit form for $A^{(3)}$ in (\ref{A3}), the identity $\gamma_\alpha\gamma_\beta=\gamma_{\alpha\beta} 
+g^{(3)}_{\alpha\beta}$,  and 
recalling that $s=1/(2gv)$, $g=1$, we precisely obtain the spinor equation (\ref{newks}). Finally, one can verify 
that the $d=4$ spinors (\ref{spinors}), with $\chi$ satisfying (\ref{3dkse}), do indeed solve (\ref{funnyKSE}).

\subsection*{The holographic free energy}

The holographic free energy of the Taub-NUT-AdS solution was discussed in \cite{Emparan:1999pm},
but of course in this latter reference there was no instanton field, which is crucial for supersymmetry. 
The calculation proceeds essentially as in section 2.5 of \cite{Martelli:2011fu}, except for the following caveat. The integrability condition for
the Killing spinor equation (\ref{funnyKSE}) gives the equations of motion following from the action 
\bea
S_{\mathrm{Euclidean}} &=& -\frac{1}{16\pi G_4}\int \diff^4x\sqrt{\det g_{\mu\nu}}\left(R + 6g^2 + F^2\right)~,
\eea
which has \emph{opposite} (relative) sign for the gauge field term compared with (\ref{4dSUGRA}) (see also \cite{Dunajski:2010uv}). This is clear from the fact that our equation 
(\ref{funnyKSE}) was obtained from the Lorentzian form of the equation by  sending $A\to \ii A$. It is therefore natural to expect that in the computation of the holographic free energy
we have to evaluate
the action $S_{\mathrm{Euclidean}}$ on shell.

Setting $g=1$ and cutting off the space at $r=R$, the bulk gravity contribution is given by
\bea
I^{\mathrm{grav}}_{\mathrm{bulk}} &=& \frac{3}{8\pi G_4}\int \diff^4  x\sqrt{\det g_{\mu\nu}}\
 = \   \frac{4 \pi \s R^3}{G_4}  - \frac{12\pi \s^3  R}{G_4}   +  \frac{8 \pi \s^4 }{G_4} ~.
\eea
Denoting by $R[\gamma]$  the scalar curvature of the boundary metric, and by $K$ the trace of its second fundamental form, 
the combined gravitational boundary terms 
\bea
I_\ct^\gr +I_\bd^\gr & = &  \frac{1}{8\pi G_4}\int \dd^3x \sqrt{\det \gamma_{\alpha\beta}}  \left(  2 + \frac{1}{2} R [\gamma]  -  K  \right)
\label{gravbdry}
\eea
have the following asymptotic expansion 
\bea
I_\ct^\gr +I_\bd^\gr & = & -\frac{4 \pi \s R^3}{G_4}  + \frac{12\pi \s^3  R}{G_4}   + \frac{4\pi\s^2(1-4\s^2)}{G_4} + {\cal O}(1/R)~,
\eea
where in particular notice there is a non-zero \emph{finite} contribution. 
The instanton action is\footnote{Notice that when $1-4s^2 \geq 0$ this term becomes negative. The calculation is however valid for any value of $s>0$.}
\bea
I^F_{\mathrm{bulk}} &=& -\frac{1}{16\pi G_4}\int \dd^4 x \sqrt{\det g_{\mu\nu}} F_{\mu\nu}F^{\mu\nu} \ =
  \ - \frac{2\pi \s^2(1-4\s^2)}{G_4}~.
\eea
Therefore the total on-shell action $S_{\mathrm{Euclidean}}$, obtained after removing the cut-off ($R \to \infty$), is given by 
\bea
I & = & I^{\mathrm{grav}}_{\mathrm{bulk}}  + I_\ct^\gr +I_\bd^\gr + I^F_{\mathrm{bulk}} \ =\ \frac{2\pi \s^2}{G_4}~.
\eea
Since the round sphere result\footnote{To recover the result for $S^2\times S^1$ boundary, one should first change coordinates back to the form 
in (\ref{intermedi}), and then set $\s=0$ there. In these coordinates, with $\tau \in [0, 2\pi]$ 
the gravitational contribution to the free energy is half that of the round sphere.}  is $\s=1/2$, we thus see that
\bea
I_{s} & =&   \frac{2\pi \s^2}{G_4} \ = \ (2\s)^2 I_{s=1/2}~, 
\eea
which  since $v=1/(2s)$ precisely agrees with the field theory result (\ref{newfree}). 

\section{A supersymmetric Eguchi-Hanson-AdS solution}
\label{bolty}

In the previous section the Taub-NUT-AdS solution  existed for both $1-4g^2\s^2\geq 0$ and 
$1-4g^2\s^2\leq 0$, 
with the sign determining whether the gauge field is real or purely imaginary, in a 
fixed choice of Wick rotation. However, in this section we consider a different 
solution which exists only when $1-4g^2\s^2\leq 0$, or equivalently $|s|\geq 1/(2g)$.
In this case the Euclidean supersymmetry equation takes the same form as the Lorentzian 
equation (\ref{LKSE}), namely
\bea
\left[ \nabla_\mu + \tfrac{1}{2} g \Gamma_\mu - \ii g A_\mu + \tfrac{\ii}{4} F_{\nu\rho} \Gamma^{\nu\rho} \Gamma_\mu \right] \epsilon &=& 0~.
\label{KSE2}
\eea
We will show that  there is a one-parameter family of  regular solutions in this class, of topology $M_4=T^*S^2$, for which there are Killing spinors solving (\ref{KSE2}).

When $|s|\geq 1/(2g)$ we may rewrite (\ref{jimllfixit}) as
\bea\label{realBPS}
P &=& -\s\sqrt{4g^2\s^2-1}~, \qquad M \ =\ Q\sqrt{4g^2\s^2-1}~,
\eea
which are now \emph{real}. Again setting  $\tau=2\s\psi$, the metric takes the form given in 
(\ref{TNAdS})
where now
\bea\label{unlikely}
\Omega(r) &=& g^2(r^2-\s^2)^2-\left[r\sqrt{4g^2s^2-1}+Q\right]^2~.
\eea
It will be useful to note that the four roots of $\Omega(r)$ in (\ref{unlikely}) are
\bea\label{roots}
\left\{\begin{array}{c} r_4 \\ r_3\end{array}\right\} &=& \frac{1}{2g}\left[\sqrt{4g^2\s^2-1}\pm \sqrt{8g^2\s^2+4gQ-1}\right]~,\nonumber\\
\left\{\begin{array}{c} r_2 \\ r_1\end{array}\right\} &=& \frac{1}{2g}\left[-\sqrt{4g^2\s^2-1}\pm \sqrt{8g^2\s^2-4gQ-1}\right]~.
\eea
Notice that these are all complex if $|s|< 1/(2g)$.
The gauge field is given by (after a suitable gauge transformation)
\bea\label{noninstanton}
A &=&  - \frac{s}{r^2-s^2}\left[2Qr+(r^2+s^2)\sqrt{4g^2\s^2-1}\right]\, \sigma_3~.
\eea
As $r\rightarrow \infty$ this tends to
\bea\label{globalpart}
A &\approx & A^{(3)} \ \equiv \  -s\sqrt{4g^2s^2-1}\sigma_3~,
\eea
which is (up to analytic continuation) what we had in the previous example (\ref{A3}).

\subsection*{Killing spinors}

Taking the same Clifford algebra and spinor conventions as the previous section, and again using the orthonormal frame 
(\ref{frame}), one can verify that the integrability condition for the Killing spinor equation (\ref{KSE2}) 
leads to the algebraic relation
\bea
\epsilon_-&=& \ii \sqrt{\frac{r-s}{r+s}}\left(\begin{array}{cc}\sqrt{\frac{(r-r_3)(r-r_4)}{(r-r_1)(r-r_2)}} & 0 
\\ 0 & \sqrt{\frac{(r-r_1)(r-r_2)}{(r-r_3)(r-r_4)}}\end{array}\right)\epsilon_+~.
\eea
Here recall that $\epsilon_\pm$ are two-component spinors, and $r_i$, $i=1,2,3,4$, are the four roots of $\Omega$ in (\ref{roots}).
Substituting into the $\mu=r$ component of (\ref{KSE2}) then leads to decoupled first order ODEs, which may be solved to give
\bea\label{spinors2}
\epsilon_+ &=& \left(\begin{array}{c}\sqrt{\frac{(r-r_1)(r-r_2)}{(r-s)}} \chi_+\\ \sqrt{\frac{(r-r_3)(r-r_4)}{(r-s)}} \chi_-\end{array}\right)~, \qquad \epsilon_- \ = \ \ii \left(\begin{array}{c} 
\sqrt{\frac{(r-r_3)(r-r_4)}{(r+s)}} \chi_+ \\ \sqrt{\frac{(r-r_1)(r-r_2)}{(r+s)}} \chi_-\end{array}\right)~,
\eea
where $\chi_\pm$ are independent of $r$. The large $r$ expansion of these is given by
\bea
\epsilon_+ &=& r^{1/2}\left[\mathbb{I}_2+\left(\frac{\s}{2}\mathbb{I}_2+\frac{1}{2g}\sqrt{4g^2\s^2-1}
\tau_3\right)r^{-1}+\mathcal{O}(r^{-2})\right]\chi~,\\
\epsilon_- &=& \ii r^{1/2}\left[\mathbb{I}_2-\left(\frac{\s}{2}\mathbb{I}_2+\frac{1}{2g}\sqrt{4g^2\s^2-1}
\tau_3\right)r^{-1}+\mathcal{O}(r^{-2})\right]\chi~,
\eea
where the two-component spinor $\chi$ is again given by (\ref{chispinor}). Notice this is the same as (\ref{spinorexpand}), up 
to analytic continuation. Again using the metric expansion and three-metric in (\ref{threemetric}), we may 
extract the coefficient of $r^{1/2}$ in (\ref{KSE2}). A very similar computation to that in the previous section 
then leads to the three-dimensional Killing spinor equation
\bea\label{3dkse2}
\nabla_\alpha^{(3)}\chi- \frac{\ii \s}{2}\gamma_\alpha \chi + \ii g A^{(3)}_\beta \gamma_{\alpha}^{\ \beta}\chi & = & 0~.
\eea
Setting $g=1$ and again identifying the squashing parameter $v=1/(2s)$, notice this is identical 
 to our original equation (\ref{newks}), but where we have replaced $A^{(3)}\rightarrow -\ii A^{(3)}$. Of course, 
 given the relative difference in Wick rotations of the gauge field in two the cases, this was 
 precisely to be expected. In fact, comparing the $A^{(3)}$ (\ref{globalpart}) in this section with its counterpart 
 (\ref{A3}) in the previous section, we see that equation (\ref{3dkse2}) is in fact \emph{identical} to (\ref{newks}), due to the 
 factor of $\ii$ difference in (\ref{globalpart}), (\ref{A3}). 

The solution to (\ref{3dkse2}) is therefore given by an appropriate analytic continuation of the solution presented in \cite{Imamura:2011wg}, 
and reads
\bea\label{3dspin}
\chi &=& \ex^{\eta\tau_3/2} \mathtt{g}^{-1}\chi_0~,
\eea
where $\mathtt{g}\in SU(2)$, $\chi_0$ is a constant two-component spinor, and 
\bea
v &=& \frac{1}{\cosh \eta}~,
\eea
where $v=1/(2s)$. In terms of Euler angles $(\psi,\theta,\varphi)$, recall that
\bea\label{su2}
\mathtt{g} &=& \left(\begin{array}{cc}\cos\frac{\theta}{2}\ex^{\ii(\psi+\varphi)/2} & \sin\frac{\theta}{2}\ex^{-\ii(\psi-\varphi)/2} 
\\ -\sin\frac{\theta}{2}\ex^{\ii(\psi-\varphi)/2} & \cos\frac{\theta}{2}\ex^{-\ii(\psi+\varphi)/2}\end{array}\right)~.
\eea

\subsection*{Regularity of the metric}

We must again consider regularity of the metric (\ref{TNAdS}). A complete metric will necessarily close off at the largest root 
$r_0$ of $\Omega(r)$, which must satisfy $r_0\geq s$. From (\ref{roots}) we see that either $r_0=r_+$ or $r_0=r_-$, where it is convenient to 
define
\bea\label{rpm}
r_+ & \equiv & r_4~, \qquad r_- \ \equiv \ r_2~.
\eea
{\it A priori} the coordinate $\psi$ must have period $2\pi/n$, for some positive integer $n$, so that 
the surfaces of constant $r$ are Lens spaces $S^3/\Z_n$. 
Assuming that $r_0>s$ is strict, then the metric (\ref{TNAdS}) will have the topology of 
a complex line bundle $M_4=\mathcal{O}(-n)\rightarrow S^2$ over $S^2$, where $r-r_0$ is the radial direction away from the zero section. 

Regularity of the metric near to the $S^2$ zero section at $r=r_0$ requires 
\bea\label{regularity}
\left|\frac{r_0^2-s^2}{s\Omega'(r_0)}\right| &=& \frac{2}{n}~.
\eea
This conditon ensures that near to $r=r_0$ the metric (\ref{TNAdS}) takes the form 
\bea
\diff s^2_4 &\approx & \diff\rho^2 + \rho^2\left[\diff \left(\frac{n\psi}{2}\right)+\frac{n}{2}\cos\theta\diff\varphi\right]^2 + (r_0^2-s^2)(\diff\theta^2+\sin^2\theta\diff\varphi^2)~,
\eea
near to $\rho=0$. Here note that $n\psi/2$ has period $2\pi$. 
Imposing (\ref{regularity}) at $r_0=r_\pm$ gives
\bea\label{Qvalues}
Q &=& Q_\pm(\s) \ \equiv \ \mp \frac{128g^4s^4-16g^2s^2-n^2}{64g^3\s^2}~.
\eea
In turn, substituting $Q=Q_\pm(s)$ into (\ref{rpm}) one then finds
\bea\label{rootschoice}
r_\pm(Q_\pm(\s)) &=& \frac{1}{8g}\left[\frac{n}{g\s}\pm 4\sqrt{4g^2\s^2-1}\right]~.
\eea
Recall that in order to have a smooth metric, we require $r_0>s$. Imposing this for $r_0=r_\pm(Q_\pm(\s))$ gives
\bea
r_\pm(Q_\pm(\s)) - s &=& \frac{1}{2g}\, f_n^\pm (2gs)~,
\eea
where the function
\bea
f_n^\pm(x) &\equiv & \frac{n}{2x}-x \pm\sqrt{x^2-1}
\eea
is required to be \emph{positive} for a smooth metric with $s=x/(2g)$. Notice here that $s\geq 1/(2g)$ implies $x\geq 1$.
It is straightforward to show that $f_n^-(x)$ is monotonic decreasing on $x\in [1,\infty)$. 
For simplicity here we will restrict our attention to $n\leq 2$.\footnote{In the first version of this paper
it was argued that $n>2$ breaks supersymmetry; however, this is incorrect.}
The analysis then splits into the cases $\{n=1\}$, $\{n=2\}$, which have a qualitatively different behaviour:

\subsubsection*{$n=1$}

It is easy to see that $f_1^\pm(x)< 0$ on $x\in [1,\infty)$, and thus the metric 
(\ref{TNAdS}) cannot be made regular in this case. Specifically, $f_1^\pm(1)=-1/2$:
since $f_1^-(x)$ is monotonic decreasing, this rules out taking $r_0=r_-(Q_-(s))$ given by (\ref{rootschoice}); 
on the other hand $f_1^+(x)$ monotonically increases to zero 
from below as $x\rightarrow\infty$, and we thus also rule out $r_0=r_+(Q_+(s))$ in (\ref{rootschoice}). 

\subsubsection*{$n=2$}

It is easy to see that $f_2^-(x)< 0$ for $x\in (1,\infty)$, 
while $f_2^+(x)>0$ on the same domain, which means we must set
\bea\label{Qset}
Q &\equiv & Q_+(s) \ = \  - \frac{(4g^2\s^2-1)(1+8g^2\s^2)}{16g^3\s^2}~,
\eea
and
\bea\label{r0}
r_0(s) & = & \frac{1}{4g}\left[\frac{1}{g\s}+ 2\sqrt{4g^2\s^2-1}\right]~,
\eea
may then be shown to be the largest root of $\Omega(r)$, for all $s\geq 1/(2g)$. 
In particular, 
this involves showing that $r_0(s)-r_-(Q_+(s))>0$  for all $s\geq 1/(2g)$, which follows since
\bea
r_0(s)-r_-(Q_+(s)) &=& \frac{1}{2g}\, h(2gs)~,
\eea
where we have defined
\bea
h(x) & \equiv & \frac{1}{x} + 2\sqrt{x^2-1}-\sqrt{4x^2-2-\frac{1}{x^2}}~.
\eea
It is a simple exercise to prove that $h(x)>0$ on $x\in (1,\infty)$.

\

After this slightly involved analysis, for $n=2$ we end up with a smooth complete metric on $M_4=T^*S^2$, given by 
(\ref{TNAdS}), (\ref{unlikely}) with $Q=Q_+(s)$ given by (\ref{Qset}),  for all $s> 1/(2g)$. 
The $S^2$ zero section is at $r=r_0(s)$ given by (\ref{r0}). The metric is thus of 
Eguchi-Hanson-AdS type, although we stress that it is \emph{not} Einstein 
for any $s>1/(2g)$. The large $r$ behaviour is again given by (\ref{approxmetric}), 
so that the conformal boundary is a squashed $S^3/\Z_2$. The $s=1/2g$ 
limit gives a round $S^3/\Z_2$  at infinity
with the bulk metric being the singular AdS$_4/\Z_2$,  albeit with a non-trivial torsion gauge field, as we shall 
see momentarily. 

It follows that another interesting difference to the Taub-NUT-AdS solution of the previous section is that 
the gauge field (\ref{noninstanton}) no longer has (anti)-self-dual field strength $F=\diff A$; moreover, 
the latter has a non-trivial flux. Indeed, 
although the gauge potential in (\ref{noninstanton}) is \emph{singular} on the 
$S^2$ at $r=r_0$,  one can easily see that the field strength $F=\diff A$ is a 
globally defined smooth two-form on our manifold. One 
 computes the period of this through the $S^2$ at $r_0(s)$ to be 
\bea\label{fluxy}
\frac{g}{2\pi}\int_{S^2} F &=& -\frac{2gs}{r_0(s)^2-s^2}\left[-2Q_+(s)r_0(s)-(r_0(s)^2+s^2)\sqrt{4g^2\s^2-1}\right]\nonumber\\
&=& 1~,
\eea
the last line simply being a remarkable identity satisfied by the largest root $r_0(s)$. Setting $g=1$, we thus 
see that we have precisely one unit of flux through the $S^2$! It follows that the gauge field $A$ is 
 a connection on the non-trivial line bundle 
$\mathcal{O}(1)\rightarrow T^*S^2$. The corresponding first Chern class $c_1=[F/2\pi]\in H^2(T^*S^2;\Z)\cong \Z$ 
is the generator of this group. Moreover, the map 
$H^2(T^*S^2;\Z)\rightarrow H^2(S^3/\Z_2;\Z)\cong\Z_2$ that restricts the gauge 
field to the conformal boundary is reduction modulo $2$. Hence
at infinity the background gauge field is more precisely given by the 
global one-form (\ref{globalpart}) \emph{plus} the flat
non-trivial Wilson line that represents the element $1 \in H^2(S^3/\Z_2;\Z)\cong H_1(S^3/\Z_2;\Z)\cong\Z_2$. 
One would be able to see this explicitly by writing the gauge field $A$ as a one-form that is locally well-defined 
 in coordinate patches, and undergoes appropriate gauge transformations between these coordinate patches. 
It follows that the gauge field at infinity is more precisely a connection on the non-trivial 
torsion line bundle over $S^3/\Z_2$.

\subsection*{The holographic free energy}

Although we will not pursue the holographic interpretation of this solution in the present paper, below we will
compute its holographic free energy using  standard formulas.
Since the gauge field here is real, the relevant action is the Euclidean action with standard signs
\bea
S_{\mathrm{Euclidean}} &=& -\frac{1}{16\pi G_4}\int \diff^4x\sqrt{\det g_{\mu\nu}}\left(R + 6g^2 - F^2\right)~.
\eea
Notice that upon taking the trace of the Einstein equation, we see that \emph{all} solutions
(supersymmetric or not) of $d=4$ gauged supergravity are metrics with  constant scalar curvature 
$R = - 12 g^2$. Using this, 
a straightforward calculation then gives for the total (bulk plus boundary) gravity part a finite result, after 
sending the cut-off $r=R\to \infty$. Namely, after setting $g=1$ we get 
\bea
I^{\mathrm{grav}}_\mathrm{tot} \  =\  I^{\mathrm{grav}}_{\mathrm{bulk}}  +I_\bd^\gr  + I_\ct^\gr & =&  \frac{\left(1-12 s^2\right) \pi }{32 G_4 s^2}\nonumber \\
& - &   \frac{ \sqrt{4 s^2-1} \left(1+4 s^2 \left(-3+8 s^2\right)\right) \pi }{16 G_4  s}~,
\eea
where we note that the contribution on the second line comes entirely from the boundary terms.
Although the gauge field is not (anti-)-self-dual, it is straightforward to compute its on-shell action, 
which is still finite, namely we get 
\bea
 I^F_{\mathrm{bulk}} &=& \frac{(1+4  s^2)\pi}{32 G_4  s^2} -\frac{\sqrt{4 s^2-1} \left(1+4 s^2 \left(1-8 s^2\right)\right) \pi }{16 G_4  s}~.
\eea
Therefore for  the total on-shell action 
we obtain
\bea
I & = & \frac{\pi}{2 G_4} + \left(s^2 - \tfrac{1}{4}\right)^{3/2}\frac{\pi }{G_4 s }~.
\label{predict}
\eea
 Notice this  makes sense for any $s>1/2$. Moreover, in the $s\to 1/2$ limit the second term vanishes and we are left with 
a result that is the same as that for the round three-sphere $S^3$. This might seem a contradiction, but in fact if we 
look back at where this result comes from, we see that in this limit 
\bea
\lim_{s\to 1/2} I^{\mathrm{grav}}_\mathrm{tot} &  = & \frac{\pi}{4 G_4}~,
\eea
which is the correct contribution expected from the (singular) AdS$_4/\Z_2$ solution with round $S^3/\Z_2$ boundary. However, we get
an equal non-zero contribution from the gauge field action
\bea
\lim_{s\to 1/2} I^F_{\mathrm{bulk}} &  =& \frac{\pi}{4 G_4}~,
\eea
despite the fact that the gauge field curvature $F\to 0$ in this limit. The calculation captures correctly the contribution from the flat torsion gauge field, which indeed cannot be 
turned off continuously since in the bulk has one unit of flux through the
 vanishing $S^2$ at the $\Z_2$ singularity.
More precisely, the complement of the singular point
has toplogy $\R_+\times S^3/\Z_2$, and the gauge field is a flat connection on the non-trivial 
torsion line bundle over this.

\section{Discussion}
\label{discussione}

In this letter we have extended the results of \cite{Martelli:2011fu}, discussing a new
class of supersymmetric solutions of $d=4$, $\mathcal{N}=2$ gauged supergravity, which in turn
uplift to solutions of eleven-dimensional supergravity. The solutions in section \ref{nuts} provide the holographic duals
to ${\cal N}=2$  supersymmetric gauge theories on the background of a biaxially squashed three-sphere and a $U(1)$ gauge field, 
whose localized partition function was recently computed in \cite{Imamura:2011wg}. In particular, as in \cite{Martelli:2011fu}, we have shown that the 
bulk metric, gauge field, and Killing spinors reduce precisely to their field theory counterparts on the boundary. 
Moreover, the holographic free energy is identical to the leading large $N$ contribution to the field theoretic free energy computed from 
the quiver matrix model. The solution is a special case of the general class of supersymmetric 
Plebanski-Demianski solutions \cite{AlonsoAlberca:2000cs}, but it differs from the solution discussed in \cite{Martelli:2011fu} in various respects. 
The graviphoton field is again an instanton, hence  the bulk metric is Einstein, but it is not now diffeomorphic to AdS$_4$. 
The results of \cite{Martelli:2011fu}, and of this letter,  suggest that the AdS/CFT correspondence is a useful setting 
for studying supersymmetric gauge theories on curved backgrounds. 

We conclude  noting that although the results presented here share a number of similarities with those in 
\cite{Chamblin:1998pz,Hawking:1998ct}, there are some crucial differences that are worth summarizing. In contrast to the solutions we have discussed,
the AdS-Taub-NUT and AdS-Taub-Bolt solutions in \cite{Chamblin:1998pz,Hawking:1998ct}  are \emph{not} supersymmetric, and moreover no 
gauge field was turned on. In addition, while those solutions have the same biaxially squashed $S^3$  boundary, the boundary of our  Eguchi-Hanson-AdS 
solution has the different topology $S^3/\Z_2$. Therefore, although we have computed the free energy for both families, it does not make 
sense to compare them along the lines of \cite{Chamblin:1998pz,Hawking:1998ct}.
It would be very interesting to understand the precise 
field theory dual interpretation of the Eguchi-Hanson-AdS solution discussed here.

\subsection*{Acknowledgments}
\noindent
We thank Jan Gutowski for a useful discussion. D.~M. is supported by an EPSRC Advanced
 Fellowship EP/D07150X/3 and J.~F.~S. by a Royal Society University Research Fellowship.

\end{document}